\documentclass[aip,jcp,twocolumn,superscriptaddress,longbibliography,10pt]{revtex4-1}
\usepackage{graphicx}
\usepackage{amssymb,amsmath,amsbsy,amsfonts,bm}
\usepackage{epsfig}
\usepackage{subfigure}
\usepackage{multirow}
\usepackage{tikz}
\usepackage{hyperref}
\usepackage{gensymb}
\usepackage{threeparttable}
\usepackage[percent]{overpic}
\usepackage{color}

\graphicspath{{Figures/}{/.}}

\begin{document}
\title{\Large Collective mechano-response dynamically tunes cell-size distributions in growing bacterial colonies}

\author{Ren\'e Wittmann}
\email{rene.wittmann@hhu.de}
\affiliation{Institut f\"ur Theoretische Physik II: Weiche Materie, Heinrich-Heine-Universit\"at D\"usseldorf, Universit\"atsstra{\ss}e 1,
D-40225 D\"usseldorf, Germany}
\author{G.\ H.\ Philipp Nguyen}
\affiliation{Institut f\"ur Theoretische Physik II: Weiche Materie, Heinrich-Heine-Universit\"at D\"usseldorf, Universit\"atsstra{\ss}e 1,
D-40225 D\"usseldorf, Germany}
\author{Hartmut L\"owen}
\affiliation{Institut f\"ur Theoretische Physik II: Weiche Materie, Heinrich-Heine-Universit\"at D\"usseldorf, Universit\"atsstra{\ss}e 1,
D-40225 D\"usseldorf, Germany}
\author{Fabian J.\ Schwarzendahl}
\email{fabian.schwarzendahl@hhu.de}
\affiliation{Institut f\"ur Theoretische Physik II: Weiche Materie, Heinrich-Heine-Universit\"at D\"usseldorf, Universit\"atsstra{\ss}e 1,
D-40225 D\"usseldorf, Germany}
\author{Anupam Sengupta}
\email{anupam.sengupta@uni.lu}
\affiliation{Physics of Living Matter Group, Department of Physics and Materials Science, University of Luxembourg, 162 A, Avenue de la Faïencerie, L-1511 Luxembourg City, Luxembourg}
\affiliation{Institute for Advanced Studies, University of Luxembourg, 2 Avenue de l’Université, L-4365, Esch-sur-Alzette, Luxembourg}
\date{\today}

\begin{abstract}\noindent
\textbf{\large Abstract}\\

Mechanical stresses stemming from environmental factors are a key determinant of cellular behavior and physiology.
Yet, the role of self-induced biomechanical stresses in growing bacterial colonies has remained largely unexplored.
Here, we demonstrate how collective mechanical forcing plays an important role in the dynamics of the cell size of growing bacteria.
We observe that the measured elongation rate of well-nourished \textit{Escherichia coli} cells decreases over time, depending on the free area around each individual, and associate this behavior with the response of the growing cells to mechanical stresses.
Via a cell-resolved model accounting for the feedback of collective forces on individual cell growth, we quantify the effect of this mechano-response on the structure and composition of growing bacterial colonies, including the local environment of each cell.
Finally, we predict that a mechano-cross-response between competing bacterial strains with distinct growth rates affects their size distributions.
\end{abstract}

\maketitle

\newpage\mbox{}\newpage\mbox{}\newpage

\noindent
\subparagraph*{\large\!\!\!\!\!\!\! Introduction}\mbox{}\\

Mechanical environment is a key determinant of behavior, physiology and functions in diverse cellular systems including microorganisms and their interactions with their host cells \cite{dufrene2020mechanomicrobiology,harper2020cell}.
 In bacterial systems, environments impose external mechanical constraints via viscous, elastic or surface forces, with far-reaching ramifications on their survival, fitness and resistance to biochemical agents, including antibiotics \cite{persat2015mechanical,genova2019mechanical}.
 A growing bacterial colony \cite{allen2018bacterialreview} presents a complex biophysical setting where an interplay of extensile (due to growth) and adhesive (cell-cell and cell-substrate) interactions engender active biomechanical constraints, which evolve as the colony expands \cite{you2018geometry,you2019monotomultilayer}.
Recently, it has been shown that such mechanical constraints give rise to active self-induced stresses within bacterial and yeast colonies, the magnitude of which depends on the local position within the colony \cite{chu2018self,alric2022macromolecular}.
Beyond a certain threshold, the response to these stresses can drive shifts in the phenotypic traits of the individual cells and
trigger critical structural changes in a colony, thereby initiating biofilm formation \cite{chu2018self,sengupta2020microbial,
dhar2022escape2d_time}.

At the level of a single cell, external mechanical influences on bacterial growth, including fluid flows and confinements, have been extensively investigated \cite{tuson2012measuring,persat2015mechanical,cesar2017fems}.
 Moreover, different models for the growth kinetics of individual bacteria \cite{amir2014cell,campos2014constant,taheri2015cell,si2019mechanistic,delgado2022influence} describe cell division events as regulated toward achieving \textit{cell-size homeostasis}.
However, owing to the small number of cells involved in these studies, dynamical effects arising from temporal and spatial biomechanical constraints, are not fully included.
Despite the crucial influence that self-induced mechanical forces seem to have on microbial behavior and physiology, currently, we lack a cell-based mechanistic model that could capture the collective interrelations and feedback between cells that spontaneously evolve at the scale of the colony.

Motivated by the gaps in our current understanding, here we report on the phenomena arising from self-induced collective mechanical stresses between the cells in expanding bacterial colonies.
While mechanical interactions have already been identified as a key ingredient to determine the structure and shape of a colony \cite{farrell2013mechanically,farrell2017mechanicalinteractionsmutants,schnyder2020control,langeslay2022},
we also describe a mechano-response of bacterial cells
that tunes the emerging distribution of cell sizes (or lengths)~\cite{puliafito2012collective} during the evolution of the growing colony.
This mechano-response
represents a cell's ability to sense local surroundings, stimuli as well as the presence of other cells \cite{cox2018bacterialmechanosensors,gordon2019bacterialmechanosensing}
and, based on this information, adapt the growth behavior to avoid overcrowding or trigger cell death \cite{podewitz2015tissue,podewitz2016interface}.
A biochemical origin could lie in the inhibited signaling under mechanical stresses~\cite{mishra2017protein,van2019mechanical}.
To mimic these effects we devise an analytically accessible model founded on statistical mechanics.
Inspired by the success of modeling various driven systems by means of rod-like particles
\cite{vicsek2012collective,baer2020SPRreview,kraikivski2006enhanced,narayan2007GNFgranularnematic,diaz2020domain,kumar2019trapping,kozhukhov2022mesoscopic},
we describe the bacteria as rigid rods of variable length \cite{allen2018bacterialreview,you2018geometry,you2019monotomultilayer}.
Our model both explains the collective self-regulation of phenotypic bacterial traits over time \cite{dhar2022escape2d_time} and unveils that collective mechanical interactions also enable bacterial populations to dynamically tune the size (or length) of single cells.

To investigate the implications of collective stresses in a controlled setting,
we perform our growth experiments on a nutrient-rich agarose substrate and put forward a microscopic first-principles model derived from dynamical density functional theory (DDFT) \cite{marconi1999DDFT,archer2004DDFT,tevrugt2020revDDFT}, which explains the central experimental observations.
Its basic ingredients are the division of a bacterium with length $2L$ into two shorter agents of length-at-birth $L$, growth with a certain elongation rate, fluctuations of the elongation rate and a mechano-response to collective interactions.
The cell-length-dependent density \cite{chauviere2012DDFT_Tumor,al2018DDFT_Tumor} in our DDFT describes the phenotemporal properties of the growing colony, i.e., the cells' length-distribution dynamics.
This allows us to gain analytic insight into the growth process, predicting the evolution towards a unique distribution of cell lengths, which reflects the stochastic nature of biological systems.
A numerical evaluation further sheds light on how the length distribution depends on the spatial position.
On top, we introduce a refined cell-based simulation tool, which reveals the detailed spatiotemporal aspects of the mechano-response to the local colony structure.
Our qualitative findings are robust with respect to implementing different mechanisms for growth or cell division \cite{amir2014cell,delgado2022influence,campos2014constant,taheri2015cell,si2019mechanistic}.
Therefore, our techniques can be applied to a broad class of biological systems,  providing a comprehensive understanding of collective biomechanical forces during population growth,
as we exemplify by investigating a mechano-cross-response between two competing bacterial strains.

\begin{figure*}
    \centering
    \includegraphics[width=0.995\textwidth]{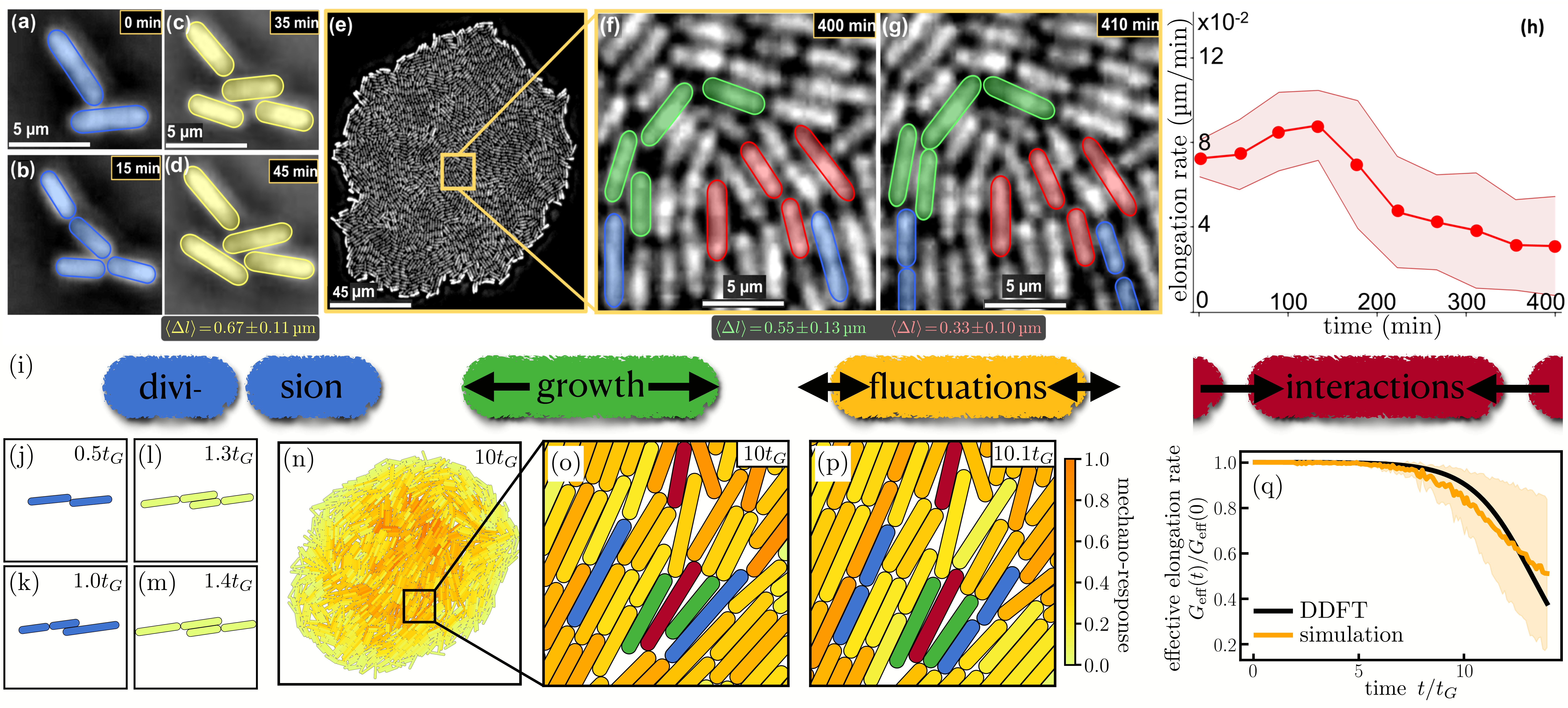}
    \caption{
    \textbf{Decreasing bacterial elongation rate in a growing colony.}
   (a-h)  Experiments on a monolayer of cells dividing every $\sim$30 minutes on a nutrient-rich agarose substrate, reaching a cell count of $\sim$4500 bacteria after 450 minutes.
       (a, b) Division of cells (hued in blue) in the early stage of the colony.
       (c, d) Growth of cells (hued in yellow) during the early phase over an interval of ten minutes. The average cell elongation rate
       (determined from the average length $\langle\Delta l\rangle$ grown in ten minutes) is $0.067\pm 0.011$ $\mu$m/min.
       (e) Micrograph of a well-developed bacterial monolayer comprising about 2600 cells.
       (f, g) Division and growth of cells over an interval of 10 minutes within the dense colony.
Dividing cells are hued in blue.
Fast and slowly growing cells are hued in green or red, respectively, as identified from a large or small free area near the ends visible in (f).
The average elongation rates resulting from this selection are $0.055\pm 0.013$ $\mu$m/min and $0.033\pm 0.01$ $\mu$m/min.
       (h) Cell elongation rate of all bacteria as a function of time (red dots) with standard deviation (red shaded area), determined from three distinct biological replicates.
       (i) Illustration of our theoretical model.
       (j-q) Model predictions, where $t_G:=L/G$ is the time between the first two divisions.
    (j-p) Snapshots of cell-based simulations reflecting the experimental images (a-f).
    The strength of the mechanical interactions on each bacterium is highlighted according to the color bar, increasing from yellow to orange.
    In the closeups of the dense colony (o,p), we also highlight selected cells showing division (blue), fast growth in a dilute region (green) and slow growth in a dense region (red).
    (q) Effective elongation rate $G_\text{eff}(t)$, defined in Eq.~\eqref{eq_Geff}, from DDFT (black), using the analytic result of Eq.~\eqref{eq_DDFTcgSOLrho},
    and simulation (orange line) with standard deviation (orange shaded area), determined from five independent simulation runs.
    The fluctuation parameter is $D=10^{-2}GL$, while the strength parameters $S=10^{-4}G/\bar{\rho}_0$ and $\tilde{S}=5\times10^3G/L^{3/2}$ of the mechano-response are chosen to obtain a comparable decay behavior in qualitative agreement with the experiment (h).
    }
    \label{figure1}
\end{figure*}

\noindent
\subparagraph*{\large\!\!\!\!\!\!\! Results}

\subparagraph*{Overview.}\vspace*{-1em}

 As compiled in Fig.~\ref{figure1}, our model is designed to capture the fundamental aspects of bacterial growth (see the Methods section and Supplementary Notes 1 and 2 for more details).
In our experiments, shown in Fig.~\ref{figure1}a-h, we observe the growth of \textit{Escherichia coli} at 30$^\circ$ C, on a nutrient-rich agarose substrate.
A single cell first divides into two cells of equal length which then continue to grow with a certain elongation rate.
At later times, as the colony becomes denser, growth and cell division progress with reduced elongation rate.
This effect is significantly more pronounced for closely packed cells whose ends are in direct contact with their neighbors than for the loosely packed cells flanked by void regions of the colony.
Overall, we observe that the average cell elongation rate is nearly halved in the grown colony, although a sufficient amount of nutrients is still available.

As demonstrated by means of cell-resolved Langevin simulations in two dimensions, shown in Fig.~\ref{figure1}j-q,
the central experimental observations are modeled by accounting for four central ingredients in the dynamics of the bacterial length (Fig.~\ref{figure1}i):
 cell division into two new individuals with length-at-birth $L$ when the maximal length $l=2L$ is reached, growth with (average) elongation rate $G$, fluctuations of this elongation rate with magnitude $D$ and collective mechanical interactions generated by the same potential as the forces and torques in configurational space (see the Methods section and Supplementary Note 1 for a precise definition).
Since these interactions have a weakening effect on the bacterial elongation rate, we speak of a mechano-response with strength quantified by the parameter $S$.

\begin{figure}
    \centering
    \includegraphics[width=0.95\columnwidth]{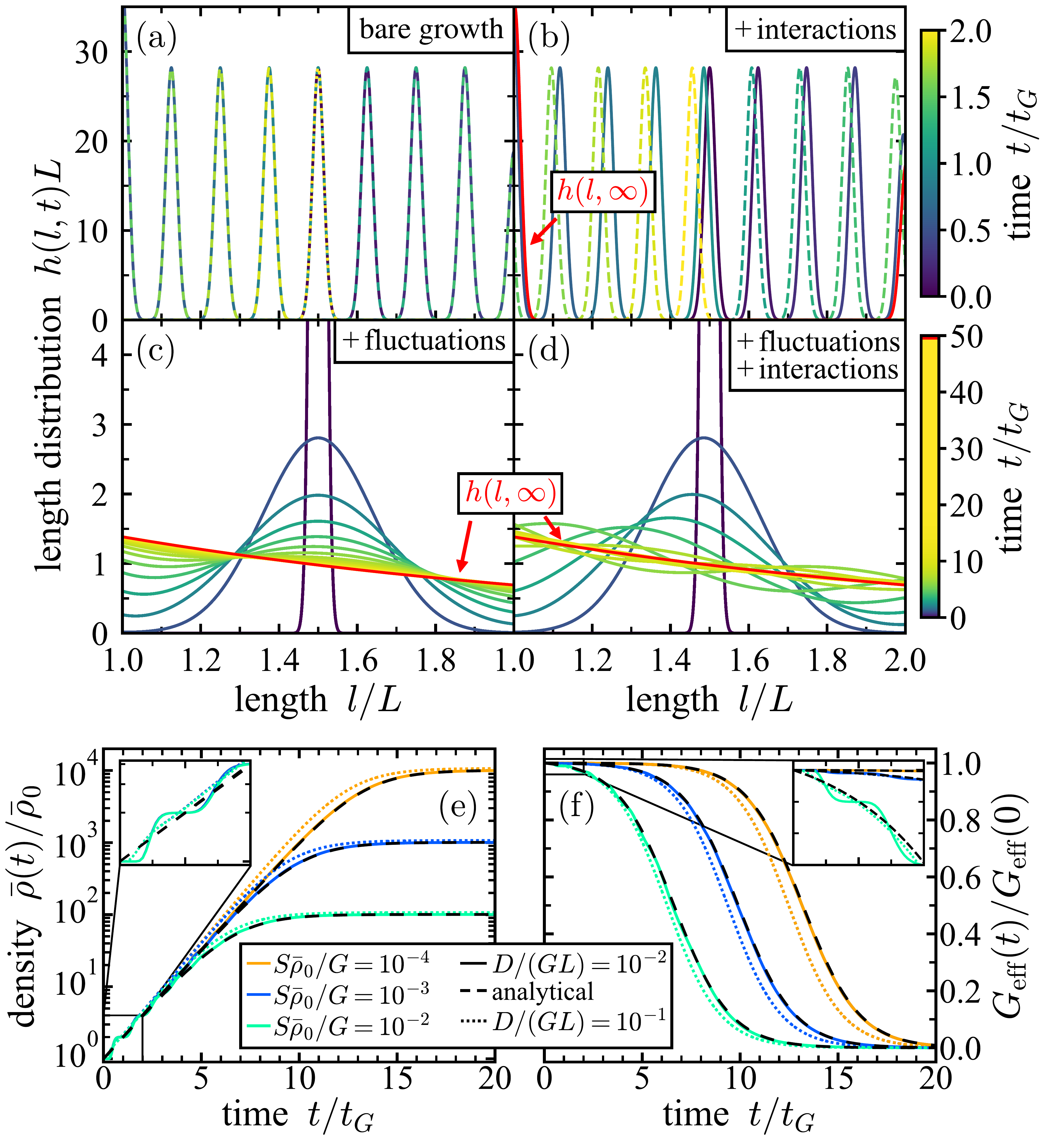}
    \caption{\textbf{Size-resolved DDFT results.} (a-d) Deterministic time evolution of the length distribution from an initial Gaussian peak centered at $l=3L/2$ with strengths $S=0$ (a,c) or $S=10^{-2}G/\bar{\rho}_0$ (b,d) of the mechano-response and growth fluctuations $D=0$ (a,b) or $D=10^{-2}GL$ (c,d).
    The time stamp of each curve is indicated by the color bar, increasing from blue to yellow.
    If a stationary distribution $h(l,\infty)$ exists (arrested peak due to mechano-response or smeared-out function due to fluctuations), it is drawn in red.
In (a,b), we show the distribution in regular time intervals of $0.2t_G$ and additionally
distinguish between the first (solid), second (dashed) and third (dotted) generation by the line style.
In (c,d) we only show each generation after multiples of $t_G$.
Animated curves are provided as Supplementary Movies 1-4.
    (e,f) Time evolution of the total density $\bar{\rho}(t)$ and the effective elongation rate $G_\text{eff}(t)$ for different parameters (as labeled). The respective analytic predictions of Eqs.~\eqref{eq_DDFTcgSOLrho} and~\eqref{eq_Geff} are drawn as black dashed lines. The insets enhance the early behavior for $t\leq2t_G$.}
 \label{figure2}
\end{figure}

\subparagraph*{Size-averaged analytical predictions.}

As the main prediction of our model, we observe in Fig.~\ref{figure1}q an effective reduction of the bacterial elongation rate over time, confirming the experimental trend in Fig.~\ref{figure1}h.
This is due to  collective interactions  ($S>0$), which slow down the growth until the colony eventually attains a stationary state when all available space is occupied.
In a first approximation, the behavior under such a mechano-response can be described by deriving a logistic growth equation for the total number density $\bar{\rho}(t)$ of all cells from our general DDFT model (see the Methods section and Supplementary Note 3).

Specifying the initial value $\bar{\rho}_0:=\bar{\rho}(0)$, we find the analytic result
 \begin{align}
\bar{\rho}(t)=\frac{R\bar{\rho}_0}{\frac{S\ln2}{L}\bar{\rho}_0+(R-\frac{S\ln2}{L}\bar{\rho}_0)e^{-Rt}}\,
  \label{eq_DDFTcgSOLrho}
\end{align}
with $R:=G\ln(2)/L+D(\ln(2)/L)^2$.
Hence, due to the mechano-response, the colony does not grow indefinitely but
approaches the maximal density $\bar{\rho}(\infty)=\frac{RL}{S\ln2}$ for $t\rightarrow\infty$.
Defining the effective elongation rate as
 \begin{align}
G_\text{eff}(t):=G+D\frac{\ln2}{L}-S\bar{\rho}(t)\,
  \label{eq_Geff}
\end{align}
and inserting $\bar{\rho}(t)$ from Eq.~\eqref{eq_DDFTcgSOLrho}, allows us to predict analytically the behavior shown in Fig.~\ref{figure1}q from DDFT, where $G_\text{eff}(t)$
  decays to zero for $t\rightarrow\infty$.
  The first two terms in Eq.~\eqref{eq_Geff} are contributions of individual cells, while the third term emerges due to collective interactions.
  Cell division merely contributes to $G_\text{eff}(t)$ by increasing the total cell count.

\subparagraph*{Size-resolved DDFT.}

While Eqs.~\eqref{eq_DDFTcgSOLrho} and~\eqref{eq_Geff} illustrate the basic concept of a mechano-response, our phenotemporal DDFT, also resolving the cell size, provides deeper insight into the growth process.
Thus, we now describe the growth dynamics of the bacterial colony by its density $\rho(l,t)$ explicitly depending on the size of the bacteria, represented here by a length $l$ (see the Methods section and Supplementary Note 2).
To make analytic progress, we restrict $l$ to the interval $l\in[L,2L]$ and, for the moment, we model cell division by the oblique boundary condition $\rho(L,t)=2\rho(2L,t)$.
We are particularly interested in the length distribution $h(l,t)=\rho(l,t)/\bar{\rho}(t)$ with $\int\mathrm{d}l\,h(l,t)=1$,
which follows from normalizing the density $\rho(l,t)$
by the total length-averaged density $\bar{\rho}(t)$.
Our DDFT results compiled in Fig.~\ref{figure2} reveal nontrivial length-distribution dynamics, where the initial condition $h_0(l):=h(l,0)$ is chosen as a sharp peak to represent a single bacterium of length $l=3L/2$.
The analytic solutions are discussed in Supplementary Note 4 and illustrated in Supplementary Figure 1.

In the absence of both fluctuations ($D=0$) and interactions ($S=0$), we observe  \textit{bare growth}, cf.\ Fig.~\ref{figure2}a, where $h(l,t)\propto 2h_0(l+\Delta l(t))+h_0(l+\Delta l(t)-L)$ is periodic in time as the horizontal offset $\Delta l(t)$ increases linearly from $0$ to $L$ within one period $t_G:=L/G$, which sets the time unit.
After each generation, denoted by an integer $n$ such that $\Delta l(t=nt_G)=0$,
the distribution resorts to its initial form $h_0(l):=h(l,0)$ and the total density $\bar{\rho}(nL/G)=2^n\bar{\rho}_0$ has doubled.

 With interactions ($S>0$ but $D=0$), we observe a \textit{decelerated growth}, cf.\ Fig.~\ref{figure2}b, where the period of $h(l,t)$ increases from generation to generation, as the effective elongation rate in Eq.~\eqref{eq_Geff} decreases over time.
 When the density $\bar{\rho}(\infty)=G/S$ reaches its stationary value in the course of the birth of a new generation, the length distribution is suddenly arrested.
 In short, the growth process comes to an end following a mechano-response to collective interactions.

  With fluctuations ($D>0$ but $S=0$), we observe \textit{disperse growth}, cf.\ Fig.~\ref{figure2}c, where the distribution smears out and approaches the unique limit $h(l,\infty)\propto2^{-l/L}$,
 without inhibiting the indefinite exponential growth of $\bar{\rho}(t)$.
 In short, the stochastic nature of bacterial growth
  results in a self-regulation of the cell length
 \cite{dhar2022escape2d_time}.

  For extreme fluctuations ($D\gg S$ and $D\gg G/\bar{\rho}_0$), this model predicts a \textit{fluctuation-driven growth} with $h(l,\infty)\propto(3L-l)$ (see Supplementary Note 4 E).

Combining the behavior from the above special cases, the full dynamics including both fluctuations ($D>0$) and interactions ($S>0$) can be understood as \textit{disperse decelerated} growth, cf.\ Fig.~\ref{figure2}d,
where the distribution $h(l,t)$ simultaneously broadens and increases its period in the course of time.
Hence, we observe in Fig.~\ref{figure2}e that the total density $\bar{\rho}(t)$ increases jerkily due to correlated division bursts in the young colony.
Later, the mechano-response sets an asymptotic threshold and the prediction from Eq.~\eqref{eq_DDFTcgSOLrho} is recovered after the cells have self-regulated their growth behavior, approaching $h(l,\infty)\propto2^{-l/L}$ due to fluctuations (other possible stationary solutions are unstable, as illustrated in Supplementary Figure 2).
 Another consequence of an initially sharp length distribution
is an effective elongation rate which depends on the instantaneous form of $h(l,t)$ (see Supplementary Note 4 D).
As shown in Fig.~\ref{figure2}f, the average effective elongation rate $G_\text{eff}(t)$, defined in Eq.~\eqref{eq_Geff}, decays earlier for both increasing mechano-response (stronger counter-force) and increasing fluctuations (faster growth).
Moreover, $G_\text{eff}(t)$ shows a jerky behavior for weak fluctuations, just like the total density.
Comparing these observations to the approximate size-averaged predictions in Eqs.~\eqref{eq_DDFTcgSOLrho} and~\eqref{eq_Geff}, also shown as a reference in Fig.~\ref{figure2}e,f for $D=10^{-2}GL$, underlines the dynamical information hidden in the length distribution.

\begin{figure*}    \centering
   \includegraphics[width=\textwidth]{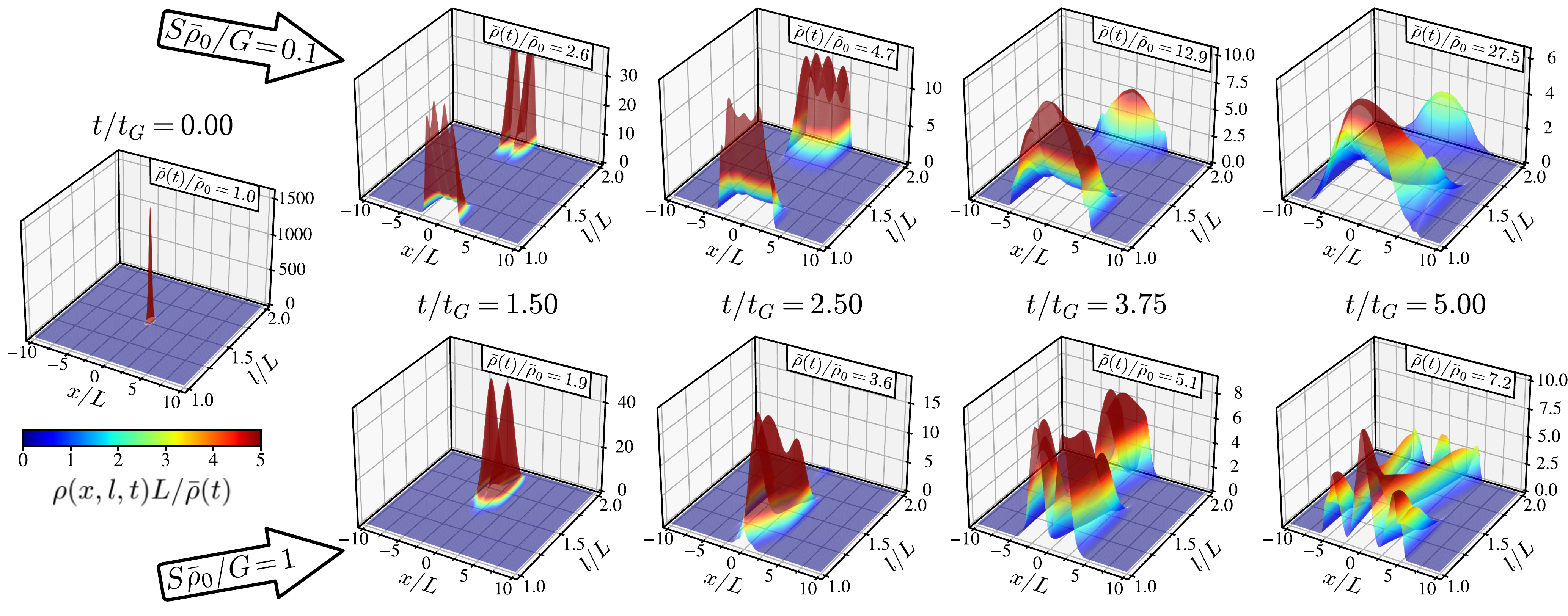}
    \caption{\textbf{Position-dependent DDFT results.} Density profile $\rho(x,l,t)$ in the early stages (time stamps are annotated in each column) of the colony evolution, normalized by the total density $\bar{\rho}(t)=\int\mathrm{d}x\int\mathrm{d}l\rho(x,l,t)/L$, as calculated with DDFT.
    Both the color bar (fixed range) and the vertical axis (adapted range for each plot) indicate the local density.
    We compare two colonies with a different mechano-response (values of $S$ are annotated in each row).
     The other parameters used in Eq.~\eqref{eq_DDFT1d} are $\gamma_x^{-1}=GL/(k_\text{B}T)$ and $D=D_x=2 \times 10^{-3}GL$.
    Additional snapshots taken at a later time and including another set of parameters are compared in Supplementary Figure 3.
    The full time evolution is provided as Supplementary Movies 5-7 for all three cases.}    \label{fig_1d}\end{figure*}

\subparagraph*{DDFT with position dependence.}

 While homogeneous systems of growing bacteria, as described with $\rho(l,t)$, can also be realized in experiments \cite{shimaya2022tilt}, we now investigate
the basic local aspects of size-resolved colony growth using our DDFT in one spatial dimension (see the Methods section and Supplementary Note 1).
Here, the density $\rho(x,l,t)$ also depends on the position $x$ in a one-dimensional channel, resembling, e.g., cells growing in a mother machine \cite{wang2010robust,yang2018analysis}.
To implement cell division, we employ a directed boundary condition, which consists of source and sink terms, representing the two daughter cells and the dividing mother cell, respectively,
and prohibits flow from short to long across the boundary (see the Methods section and Supplementary Note 5).

In general, our model describes a decrease of the local cell elongation rate in the regions with high density according to the stronger local mechano-response.
To illustrate this spatial dependence, we depict the normalized density in the early stages of the colony evolution in Fig.~\ref{fig_1d}.
Initially, we observe individual density peaks, each representing a cell, whose number doubles after the birth of the next generation.
The effect of an increasing mechano-response is to slow down the overall colony growth, as predicted in Fig.~\ref{figure2}, but here also on a local level.
For example, in the bottom-right snapshot of Fig.~\ref{fig_1d}, the outer peaks represent cells one generation ahead of those in the center, where the local density is much higher.
For a weaker mechano-response (top panel of Fig.~\ref{fig_1d}) this effect is less pronounced and we observe that the peaks gradually merge due to growth fluctuations and spatial diffusion.
Moreover, for a stronger spatial mechano-repulsion, i.e., lower substrate friction, the peaks are pushed further apart from each other (see Supplementary Figure 3), which reduces the local strength of the mechano-response, such that the total cell count increases faster.
Eventually, a smooth colony structure is approached which is characterized by having a higher density and a larger percentage of shorter cells in the center than in the periphery.

\subparagraph*{Cell-based simulations.}

To corroborate the predictions of our probabilistic DDFT and better understand the repercussions on the individual growth dynamics, we also perform cell-resolved Langevin simulations in two spatial dimensions.
As detailed in the Methods section, we consider rod-like bacteria that interact through Hertzian repulsion.
Specifically, the Langevin equations for positions and orientations are coupled to the stochastic dynamics of the cell lengths, including a response term with respect to the same Hertzian overlap potential.

\begin{figure}
    \centering
    \includegraphics[width=1.0\columnwidth]{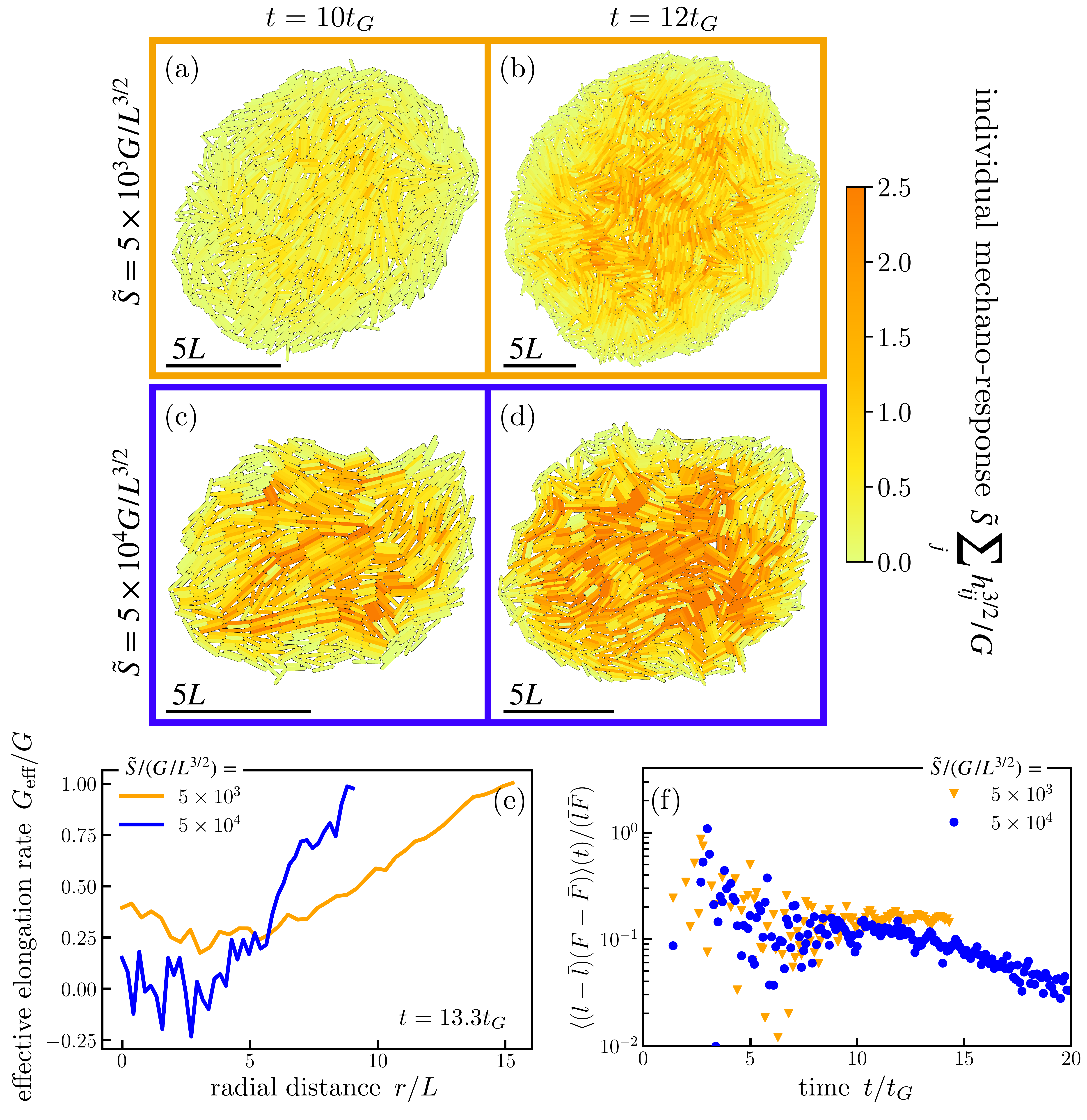}
    \caption{\textbf{Cell-based simulation results.}
     We show results for two colonies (averaged over five simulation runs) with the same growth fluctuations $D=10^{-2}GL$ but different strengths $\tilde{S}=5\times10^3G/L^{3/2}$ (orange color, as in Fig.~\ref{figure1}j-q) and $\tilde{S}=5\times10^4G/L^{3/2}$ (blue color) of the mechano-response.
     The corresponding standard deviation is of the same magnitude as in Fig.~\ref{figure1}q.
    (a-d) Snapshots of the two colonies (distinguished by the frame color), at two times $t=10t_G$ (a,c) and $t=12t_G$ (b,d).
    For comparison a scale bar of length $5L$ is drawn in each case.
         The color code indicates the mechano-response of individual bacteria, which is proportional to the overlap with all neighboring cells (as in Fig.~\ref{figure1}n but with a different range of the color bar).
    (e) Local effective elongation rate as a function of the radial distance $r$ from the colony center of mass (solid lines).
    (f) Normalized length-force correlation in both colonies at different times (symbols).
    }
    \label{fig_SIM}
\end{figure}

The effective elongation rate $G_\text{eff}(t)$ averaged over the whole colony is shown in Fig.~\ref{figure1}q.
As in the experiment (Fig.~\ref{figure1}h), $G_\text{eff}(t)$ decreases at later times but does not exponentially approach zero since the continuously growing periphery of the colony always remains sufficiently dilute.
This result is illustrated in more detail by the different snapshots compiled in Fig.~\ref{fig_SIM}a-d.
Comparing the colony size (different scale bars) after certain times, it becomes apparent that the colony of bacteria with a stronger overall mechano-response (larger $\tilde{S}$) grows slower.

We also resolve by the color code in Fig.~\ref{fig_SIM}a-d the local mechano-response of individual cells, which is apparently stronger in the center of the colony than in the periphery.
As this local quantity scales with the overall mechano-response, the
difference between the typical forces in these two regions is more significant for larger $\tilde{S}$.
Our qualitative observations are confirmed in Fig.~\ref{fig_SIM}e by measuring the local elongation rate, which is steeper for larger $\tilde{S}$ (also notice the smaller colony radius after the same amount of time) and generally increases from the colony center to the periphery, following the decrease of local density (see also the DDFT results in Fig.~\ref{fig_1d}).
The correlation depicted in Fig.~\ref{fig_SIM}f further reveals that the mechanical force on longer bacteria is typically stronger than on shorter bacteria throughout the colony evolution.

Moreover, as illustrated in Fig.~\ref{figure1}o-p, the local force acting on a bacterium in a dense colony is smaller when it is surrounded by voids than when its ends are in close contact with its neighbors, which also feeds back on the individual elongation rate, as measured experimentally in Fig.~\ref{figure1}f-g.

\begin{figure}
    \centering
    \includegraphics[width=1.0\columnwidth]{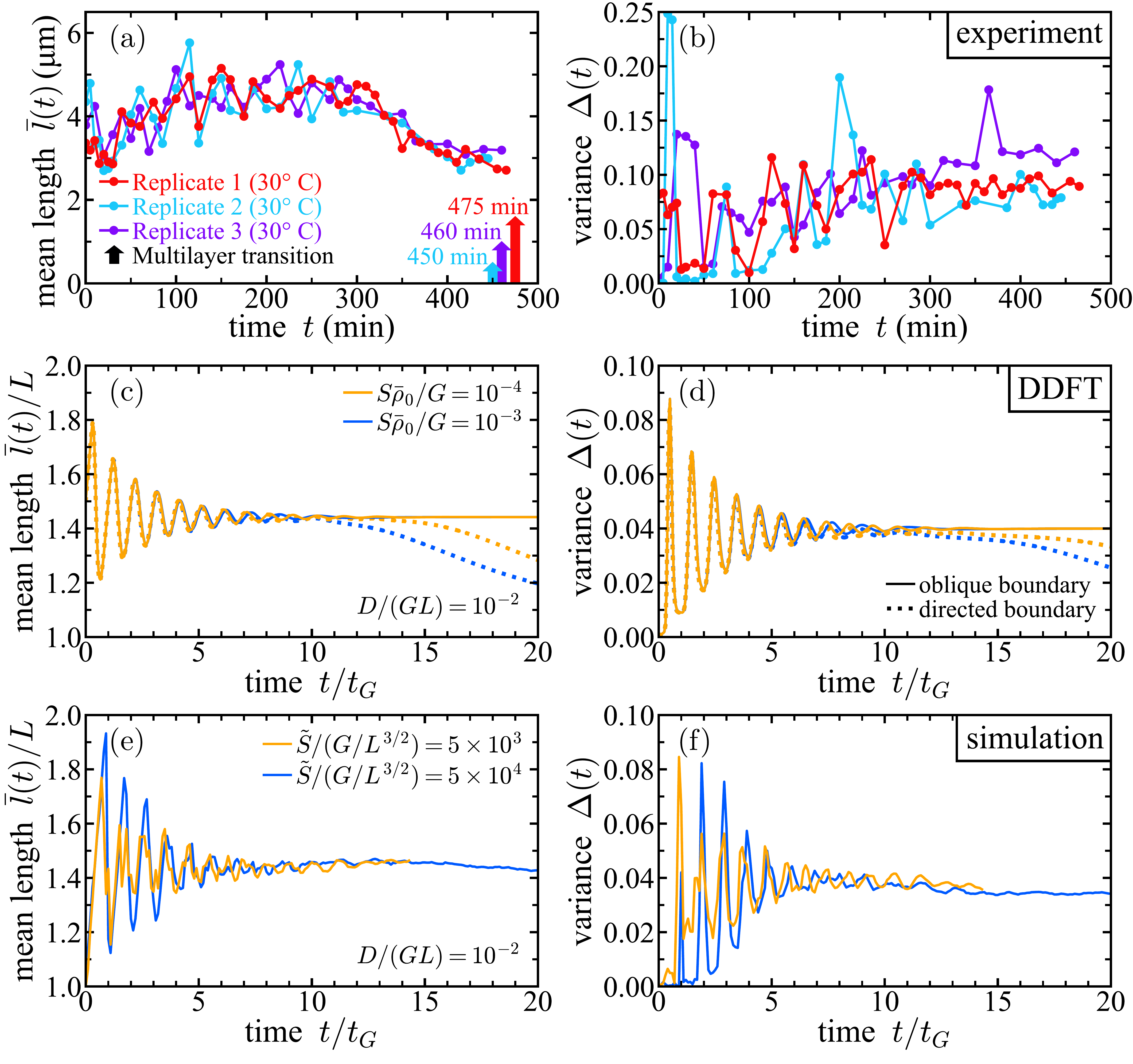}
    \caption{\textbf{Moments of the length distribution.}
    (a,b) Experiments using three biological replicates under similar growth conditions at a temperature of 30\textdegree{}C. Data are shown until the mono-to-multilayer transition~\cite{you2019monotomultilayer} (upward arrows).
   (c,d) Deterministic DDFT predictions with both oblique (solid lines)
    and directed (dotted lines) boundary conditions (see Eqs.~\ref{eq_rhodivision1dHOM} and \ref{eq_directedboundary} in the Methods section).
   (e,f) cell-based simulations.
      The simulation parameters are the same as in Fig.~\ref{fig_SIM} and chosen to be comparable to those in DDFT, cf.\ Fig.~\ref{figure1}q.
      The corresponding standard deviation is of the same magnitude as in Fig.~\ref{figure1}q.
   Blue lines in (c-f) indicate that the mechano-response is increased by a factor of ten compared to the orange lines.
    For our three different methods, we compare the mean cell length $\bar{l}(t)$ (a,c,e) and the normalized variance $\Delta(t)$ of the cell length (b,d,f).}
    \label{figure4}
\end{figure}

\subparagraph*{Length-distribution dynamics.}

To gain further insight into the length-distribution dynamics, we compare in Fig.~\ref{figure4} the experimental results for the mean length $\bar{l}(t):=\int\mathrm{d}l\,l\,h(l,t)$ and the normalized variance $\Delta(t):=\int\mathrm{d}l\,(l-\bar{l}(t))^2\,h(l,t)/(\bar{l}(t))^2$ to our model predictions.
After early fluctuations owed to the synchronous cell cycles in the early generations, the experimental data show a clear trend that the bacteria become shorter in the dense colony,
while the variance begins to plateau under the experimental growth conditions considered here.
Accordingly, the moments predicted by DDFT oscillate around their stationary values with a decreasing amplitude.
Recall that the version of our DDFT used here formally describes a homogeneous system with the cell lengths restricted to a fixed interval.

The qualitative experimental observations for later times are better captured by
utilizing the directed boundary condition introduced for our position-resolved DDFT, as it
prohibits flow from short to long (see the Methods section and Supplementary Note 5).
This refined implementation allows us to effectively predict the expected decrease of $\bar{l}(t)$ within our size-resolved DDFT.
As shown in Supplementary Figures 4 and 5,
the colony enters an additional fluctuation-driven regime after it has grown sufficiently dense,
which is characterized by a gradual approach to a distinct stationary solution,
where shorter bacteria balance the higher cell count (the smaller variance is due to our assumption of a restricted length interval).
 We also find consistent results from our cell-based simulations of a freely growing colony, where the bacteria in the dense region can effectively respond by shrinking below the length-at-birth~$L$ (see Supplementary Figure 6).
 A detailed comparison between our approaches in view of the observations in Fig.~\ref{figure4} can be found at the end of the Methods section.

\begin{figure*}[t]
    \centering
    \includegraphics[width=0.80\textwidth]{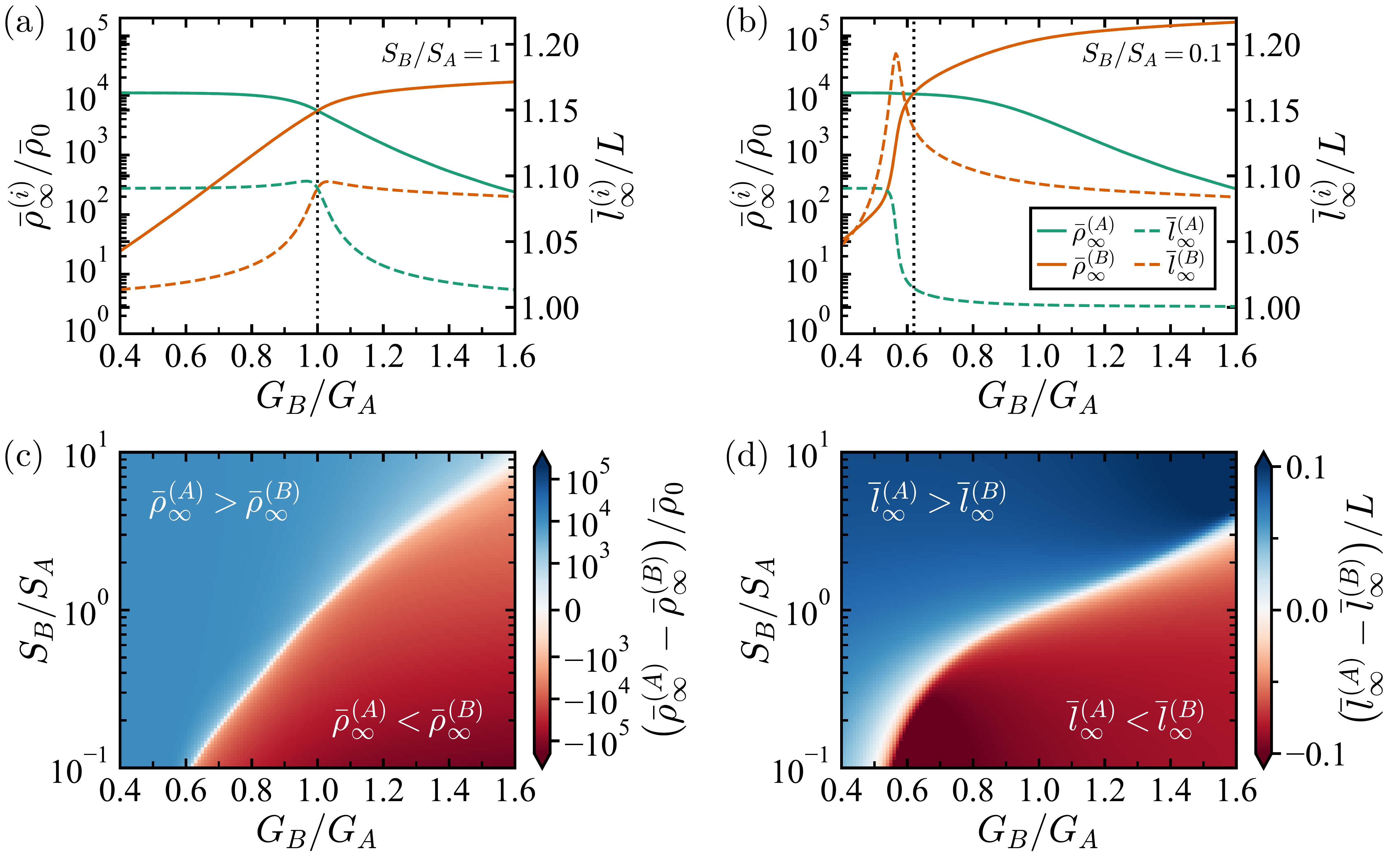}
    \caption{\textbf{Stationary properties of two competing bacterial strains.}
    We compare DDFT results of binary mixtures
        with different elongation rates $G_B$ and strength $S_B$ of the mechano-response of species $B$, but with fixed values $G_A=G$ as a scaling factor and $S_A=10^{-4}G/\bar{\rho}_0$ for species $A$ and the same growth fluctuations $D=10^{-2}GL$ of both species.
    (a,b) Stationary total densities $\bar{\rho}_\infty^{(i)}$ (solid lines) and mean lengths $\bar{l}_\infty^{(i)}$ (dashed lines) of the two strains $i=A$ (green) and $i=B$ (orange) as a function of $G_B$ for two values of $S_B$ (as labeled). The vertical dotted line indicates the value $G_B=G_\text{th}$ at which $\bar{\rho}_\infty^{(A)}=\bar{\rho}_\infty^{(B)}$ in each case.
    (c) State diagram showing the difference in the total density. As labeled, the color indicates the dominance region of species $A$ over $B$ (blue) or vice versa (red).
    The white color corresponds to $G_B=G_\text{th}$ at which a transition happens for a certain $S_B$.
    (d) State diagram showing the difference of the mean lengths in the same fashion.
    }
    \label{fig_mixture}
\end{figure*}

\subparagraph*{Mechano-cross-response of competing strains.}

When evolving on the same substrate, two different bacterial strains compete for the available resources, including the free space in the colony.
Therefore, the growth of each individual is not only regulated by the interaction with its own kind but also with the other species.
To investigate this mechano-cross-response, we consider two interacting species $A$ and $B$ with different characteristic growth properties \cite{giometto2018physical}.
For such a system, our multicomponent DDFT (see the Methods section) with directed boundary conditions predicts distinct length distributions in the fully grown homogeneous colony, whose general form is related analytically to the input parameters and final densities in Supplementary Note 6.
We can thus use this stationary information to infer the growth properties of the two competing bacterial strains.

To understand the ultimate colony composition, we compare in Fig.~\ref{fig_mixture} the stationary total densities $\bar{\rho}_\infty^{(i)}:=\bar{\rho}^{(i)}(\infty)$ and mean lengths $\bar{l}_\infty^{(i)}:=\bar{l}^{(i)}(\infty)$ with $i=A,B$ for different parameters of our model.
For equal strengths of the mechano-response ($S_A=S_B$, Fig.~\ref{fig_mixture}a), the faster-growing strain also ends up with a larger total number of cells and thus dominates the colony.
Due to the mechano-cross-response, the other strain ends up with a larger percentage of shorter cells.
Setting now, say, $S_A>S_B$ (Fig.~\ref{fig_mixture}b), a dominance of species $B$ over species $A$ is even possible if the latter has a larger elongation rate $G_A>G_B$, as long as $G_B>G_\text{th}$ exceeds an interaction-dependent threshold $G_\text{th}$ set by the condition $\bar{\rho}_\infty^{(A)}=\bar{\rho}_\infty^{(B)}$.
Slightly below this threshold, the shape of the stationary length distributions changes drastically for small variations of $G_B\lesssim G_\text{th}$.
This behavior of $\bar{\rho}_\infty^{(i)}$ and $\bar{l}_\infty^{(i)}$ is illustrated for a broader range of interaction parameters in the state diagrams in Fig.~\ref{fig_mixture}c-d.
Moreover, as exemplified in Supplementary Figure 7 and discussed in Supplementary Note~6, we observe for $G_B\gtrsim G_\text{th}$ a dynamical crossover in the total densities, $\bar{\rho}^{(A)}(t)>\bar{\rho}^{(B)}(t)$ for $t<t_\text{c}$ but $\bar{\rho}^{(B)}(t)>\bar{\rho}^{(A)}(t)$ for $t>t_\text{c}$ with the crossover time $t_\text{c}$,
and an adiabatic regime in which the faster growing but more mechano-responsive species $A$ follows a sequence of quasi-stationary length distributions after its total density has saturated.

The experimental investigation of such bacterial mixtures, together with a more comprehensive theoretical study that allows for a possible extinction of the dominated species, shall shed further light on the cross-talk and interfaces between distinct active growth processes in future work.

\subparagraph*{\large\!\!\!\!\!\!\! Discussion} \mbox{}

We have introduced a DDFT model for growing bacterial colonies from which we predicted analytical length distributions and drew parallels to dynamical observations in experiments and Langevin simulations.
Despite its simplicity, our model captures the basic features of collective interactions among the cells in \textit{in vitro} experiments ruling out possible effects of nutrient limitation.
This agreement is demonstrated by investigating the (local) reduction of the elongation rate depending on the bacterial mechano-response.
For a mixture of competing bacterial strains, our model suggests that a mechano-cross-response between the two species affects the (dynamical) length distributions in a nontrivial way.
Our general theoretical description relates these effects to a microscopic interaction potential and, therefore, it both contributes to the understanding of collective effects in models of growing bacterial colonies and elucidates the consequences of biomechanical forces for the evolution of living samples.
 While a comprehensive description of biological systems surely requires the inclusion of additional effects (as discussed below), our length-resolved tools are potentially of inherent theoretical and mathematical interest in their own right, for example regarding the interplay between cell length and topological defects.

The utility of our model can be exemplified by contemplating the onset of the mono-to-multilayer transition~\cite{you2019monotomultilayer,dhar2022escape2d_time,chu2018self}  (or verticalization~\cite{beroz2018verticalization,nijjer2021mechanical,shimaya2022tilt}), i.e., when individual bacteria evade crowded regions by escaping into the third dimension to form additional layers.
 Such structural transitions are triggered as a consequence of the self-imposed mechanical stresses that build up in a growing bacterial colony \cite{you2019monotomultilayer,dhar2022escape2d_time}.
Accordingly, our model allows us to determine directly the local in-plane mechanical force opposing the growth of each cell, cf.\ Fig.~\ref{fig_SIM}a-d, which can be compared to an appropriately chosen threshold.
In our experiments, the transition happens after about 450 minutes, which roughly corresponds to $t\!\approx\!13t_G$ for the model parameters chosen accordingly in Fig.~\ref{figure1}q.
By then the peak pressure within the colony has reached values around 10 kPa~\cite{you2019monotomultilayer}, which is consistent with the self-imposed physical pressures measured by Chu et al.~\cite{chu2018self}.
A more quantitative comparison in future work should also take into account the role of  substrate friction \cite{podewitz2016interface}.

To provide a more accurate description of specific internal processes in individual cells and give a broader account of biodiversity, our model can be readily extended in various directions.
\textit{Escherichia coli} typically exhibit exponential growth \cite{amir2014cell,delgado2022influence,dhar2022escape2d_time},
which can be modeled on a cellular level \cite{mukherjee2023cell},
and cell division occurs after adding a specific length \cite{campos2014constant,taheri2015cell,si2019mechanistic}.
These processes are intrinsically related to a cell cycle \cite{malmi2018cell,banwarth2020quantifying,colin2021twoprocessescelldivision,li2021role,li2022competition}.
One may further include death events \cite{allocati2015communitydeath,ghosh2017morphodynamicsdeath,yanni2019drivers,podewitz2015tissue,podewitz2016interface} or describe division into multiple daughter cells \cite{tse2012increased}
and then consider more general mixtures of cells with different physiological properties \cite{frost2018cooperationantibiotics,li2022competition}, also allowing for phenotype switching \cite{balaban2004bacterial,kussell2005phenotypic} or mutation \cite{wrande2008accumulationmutants,farrell2017mechanicalinteractionsmutants,hashuel2019agingmutants}.
Finally, we stress that our model, as employed here, is formally not limited to the interpretation of mimicking mechanical interactions.
An alternative application of our equations would be to conceive an interaction potential differing from that describing spatial repulsion, such that it \textit{effectively} incorporates other effects limiting cell growth, such as biochemical signaling~\cite{mishra2017protein,van2019mechanical} or nutrient depletion~\cite{hornung2018nutrientlimitedgrowth}.

Further perspectives on the spatiotemporal implications of a heterogeneous local length distribution open up when additionally resolving positions and orientations.
Both our generalized DDFT (see the Methods section) and our cell-based simulations allow us to include and compare the mechano-response with respect to different externally applied forces affecting the growth process on the single-cell level.
For example, the bacteria can be immersed in a stiff gel \cite{tuson2012measuring}, grown on rough surfaces \cite{yoda2014effect,zheng2021implication} or spatially confined \cite{volfson2008biomechanical,podewitz2015tissue,you2021confinement}.
It is also possible to investigate the detailed colony structure \cite{zhang2010collective,liu2015metaboliccodependencebiofilms,ghosh2017morphodynamicsdeath,beraghosh2022mechanisticbiofilmcoexistence},
topological defect dynamics \cite{doostmohammadi2016defect,dell2018growingcolonyactivenematic,nordemann2020defect,sengupta2020microbial,dhar2022escape2d_time,shimaya2022tilt},
emerging smectic order \cite{boyer2011buckling,wittmann2021annulus,monderkamp2021topology,xia2021structural,paget2022tensorsmectic,langeslay2022},
growth within porous media \cite{ingham1998transportinporousmediabook,or2007porousreview}
or the explicit onset of three-dimensional growth \cite{beroz2018verticalization,nijjer2021mechanical,dhar2022escape2d_time,drescher2016architecturaltransition3dcellresolution,warren2019spatiotemporal}.
Moreover, attractions through pili bonds \cite{ponisch2018pili,kuan2021continuum}, or
the motility of individual bacteria \cite{ni2020growth,sarkar2021minimal,breoni2022cycle} can be modeled.
 In view of a comprehensive biological picture, one could study
interactions and competition between multiple colonies or cell strains \cite{liu2016interspecific,ponisch2018pili,kuan2021continuum,
basaran2022inward}
alongside their cross-talk with other spatially distributed agents \cite{nadell2016spatial},
such as signaling molecules \cite{mishra2017protein,van2019mechanical}
nutrients \cite{ghosh2015biofilm,hornung2018nutrientlimitedgrowth},
antibiotics \cite{fridman2014antibiotics,frost2018cooperationantibiotics},
parasites like bacteriophages \cite{kauffman2022phages}
or a secreted extracellular matrix mediating biofilm formation \cite{ghosh2015biofilm,beraghosh2022mechanisticbiofilmcoexistence}.

Finally, our DDFT equations allow for a systematic derivation \cite{elder2007PCF_DDFT,vanteeffelen2009derivationPFC_DDFT} of phase-field crystal models \cite{elder2002PFC} to recover hydrodynamic field equations~\cite{dell2018growingcolonyactivenematic,doostmohammadi2016defect,you2018geometry}
 and explicitly incorporate aspects related to bacterial length.
  Exploring the relation to active nematics \cite{doostmohammadi2018activenematics,copenhagen2021topological} constitutes a possible direction for future work.

\newpage

\subparagraph*{\large\!\!\!\!\!\!\! Methods}

  \subparagraph*{General DDFT for growing bacterial colonies}\vspace*{-1em}

  We propose a dynamical density functional theory (DDFT) to model growing bacterial colonies through a time-dependent
density $\rho(\mathbf{r},\mathbf{p},l,t)$, which resolves the spatial position $\mathbf{r}$, orientation $\mathbf{p}$ and a size parameter $l$, which here represents the cell length.
In its most general form, the DDFT reads
  \begin{align}
\frac{\partial \rho}{\partial t}&=
-\boldsymbol{\nabla}_\mathbf{r}\cdot \mathbf{J}_\mathbf{r}
-\hat{\mathcal{R}}_\mathbf{p}\cdot \mathbf{J}_\mathbf{p}
 -\frac{\partial}{\partial l}J_l\cr
 &\ \ \ \ \,
 +\delta(l-L)J_l\left(\mathbf{r}+\frac{L}{2}\mathbf{p},\mathbf{p},2L,t\right)\cr
 &\ \ \ \ \,
 +\delta(l-L)J_l\left(\mathbf{r}-\frac{L}{2}\mathbf{p},\mathbf{p},2L,t\right)\cr
 &\ \ \ \ \,
 -\delta(l-2L)J_l\left(\mathbf{r},\mathbf{p},2L,t\right)\,,
 \label{eq_DDFTgcCONT_BC}
\end{align}
with the currents $J_\mathbf{r}(\mathbf{r},\mathbf{p},l,t)$, $J_\mathbf{p}(\mathbf{r},\mathbf{p},l,t)$ and $J_l(\mathbf{r},\mathbf{p},l,t)$ in positional, orientational and length space, respectively.
The former two terms are given in their standard DDFT form as \cite{marconi1999DDFT,archer2004DDFT,tevrugt2020revDDFT}
\begin{align}\label{eq_currents}
\mathbf{J}_\mathbf{r} &=  - D_\mathbf{r} \boldsymbol{\nabla}_\mathbf{r} \rho - \gamma^{-1}_\mathbf{r}\, \rho \, \boldsymbol{\nabla}_\mathbf{r} \left(\frac{\delta\mathcal{F}_\text{ex}\left[\rho\right]}{\delta\rho}+ V_\text{ext}\right)\,,\cr
\mathbf{J}_\mathbf{p} &=  - D_\mathbf{p} \hat{\mathcal{R}}_\mathbf{p} \rho - \gamma^{-1}_\mathbf{p}\, \rho\, \hat{\mathcal{R}}_\mathbf{p} \left(\frac{\delta\mathcal{F}_\text{ex}\left[\rho\right]}{\delta\rho}
+ V_\text{ext}\right)\,
\end{align}
with the diffusion coefficients $D_\mathbf{r}$ and $D_\mathbf{p}$,
the friction coefficients $\gamma_\mathbf{r}$ and $\gamma_\mathbf{p}$
and the derivative operators $\boldsymbol{\nabla}_\mathbf{r}$ and  $\hat{\mathcal{R}}_\mathbf{p}=\mathbf{p}\times\boldsymbol{\nabla}_\mathbf{p}$ in positional and rotational space, respectively.
The internal interactions between the cells and interactions with an externally imposed field are described by the excess part $\mathcal{F}_\text{ex}[\rho]$ of the free energy and an external potential $V_\text{ext}(\mathbf{r},\mathbf{p},l,t)$, respectively \cite{Evans1979}.

As the central ingredient of our model, the length current
  \begin{equation}
J_l=G\,\rho-D\,\frac{\partial\rho}{\partial l}-\frac{S}{k_\text{B}T}\,\rho\,\frac{\partial}{\partial l}\left(
 {\frac{\delta\mathcal{F}_\text{ex}[\rho]}{\delta\rho}}+  V_\text{ext}
\right)\,
\label{eq_DDFTlllterms}
\end{equation}
drives the length-dependent changes of the density, where the thermal energy $k_\text{B}T$ is used as a scaling factor.
The first term describes cell growth according to the growth function
$G(\mathbf{r},\mathbf{p},l,t)$ and thus drives the system out of equilibrium, while the remaining terms have a similar form as the currents in Eq.~\eqref{eq_currents} but possess a slightly different interpretation.
The term $\propto D$ is of diffusive nature and describes fluctuations of the growth function, while the terms $\propto S$ describe the response of the cell growth to internal and external interactions, where $S$ is related to the substrate friction.
Finally, the source and sink terms in Eq.~\eqref{eq_DDFTgcCONT_BC} describe cell division after a length $L(t)$ is reached.
A more detailed introduction to Eq.~\eqref{eq_DDFTgcCONT_BC} and the framework of DDFT in general can be found in Supplementary Note 1.

\subparagraph*{Overview of related approaches}

In Eq.~\eqref{eq_DDFTgcCONT_BC}  we have presented the most general form of our basic model in the language of DDFT
which we can, in principle, even further extend in several directions along the lines of the Discussion section.
What is left to be specified are the explicit interactions between the cells.
Instead of evaluating the full multidimensional DDFT, we focus in the main text on different approaches based on this model, which are further described in the remaining paragraphs of this Methods section.

Specifically, we provide in Eqs.~\eqref{eq_LEr},~\eqref{eq_LEt} and~\eqref{eq_LangevinSI} a set of stochastic Langevin equations in two dimensions, which are formally equivalent to our DDFT in Eq.~\eqref{eq_DDFTgcCONT_BC} and allow for a cell-resolved investigation of the mechano-response.
On the DDFT side, we consider various versions of Eq.~\eqref{eq_DDFTgcCONT_BC} focusing on different aspects.
First, we introduce a one-dimensional version of our DDFT in Eqs.~\eqref{eq_current1d} and~\eqref{eq_DDFT1d} to illustrate the positional dependence of the evolving length distribution.
Second, we derive a homogeneous size-resolved DDFT in Eqs.~\eqref{eq_DDFTgcCONT_BC0d} and~\eqref{eq_Jll}, which formally assumes a well-mixed system and can be analytically investigated.
Third, we also generalize this size-resolved DDFT by Eqs.~\eqref{eq_DDFTgcCONT_BC0dmix} and~\eqref{eq_Jllmix} to a version valid for multiple bacterial species.
Fourth, we demonstrate that a logistic growth equation, Eq.~\eqref{eq_DDFTcg} can be recovered upon further averaging our homogeneous DDFT over the cell size.
After presenting details on our experimental system, we conclude the Methods section with a discussion of how the different versions of our model are related and which experimental aspects we intend to describe.

\subparagraph*{Cell-based Langevin simulations}

In the particle-based approach to modeling growing bacterial colonies, the cells are considered as rigid rods.
Their positions $\bm{r}_{i}$, orientations $\theta_{i}$ and lengths $l_i$ (of their long axis) evolve in time according to coupled Langevin equations.
Here $i=1,\ldots, N$ is the cell index, where the total number $N(t)$ of bacteria may increase after each time step due to cell division.
The short axis of each rod is kept fixed with length $d_0$.

The position $\bm{r}_i$ of rod $i$ evolves according to
\begin{align}
  &  \frac{\mathrm{d}
  \bm{r}_{i}}{\mathrm{d} t}=\frac{1}{\gamma l_{i}} \sum_{j} \bm{F}_{i j}\,,
  \label{eq_LEr}
\end{align}
where $\gamma$ is the friction coefficient and $\bm{F}_{i j}$ are steric forces stemming from the interactions with other rods.
Further, the orientation of the rod is measured by the angle $\theta_i$ with respect to the $x$-axis in a Cartesian coordinate system. The dynamics of the angles are given by
  \begin{align}
&\frac{\mathrm{d} \theta_{i}}{\mathrm{d} t}=\frac{12}{\gamma l_{i}^{3}} \sum_{j}\left(\bm{r}_{i j} \times \bm{F}_{i j}\right) \cdot \bm{e}_z\,,
 \label{eq_LEt}
\end{align}
where $\bm{r}_{ij}= \bm{r}_i-\bm{r}_j$ is the distance vector between particles $i$ and $j$ and $\bm{e}_z$ is the vector perpendicular to the rods' plane of motion.
The forces between rods are calculated by a Hertzian repulsion
\begin{align}
    \bm{F}_{i j}=F_0 d_{0}^{1 / 2} h_{i j}^{3 / 2} \bm{n}_{i j}\,,
    \label{eq_Hertz}
\end{align}
where $h_{ij}$ is the overlap of rods $i$ and $j$, $F_0$ is the strength of the force, and $\bm{n}_{ij}$ is the vector normal to the closest point of contact of the particles.

In the same spirit, we now allow the length of a rod $i$ to evolve as
\begin{equation}
 \frac{\mathrm{d} l_i}{\mathrm{d}t}= G+ \sqrt{2D} \xi_i  - \tilde{S} \sum_j h_{i j}^{3 / 2}
 \label{eq_LangevinSI}
\end{equation}
with the constant elongation rate $G$, a white noise $\xi_i$ of unit variance accounting for fluctuations of magnitude $D$ and mechanical interactions mediated by the overlap between particles $h_{ij}$.
The parameter $\tilde{S}$ quantifies the strength of this mechano-response and takes the role of an inverse friction coefficient.
Here, we have absorbed the other parameters from Eq.~\eqref{eq_Hertz}, such that
$\tilde{S}$ is formally different from $S$ in Eq.~\eqref{eq_DDFT1dHOMexplicit} given the different nature of interactions considered.
 When the length $l_i$ of a cell exceeds the value $2L$ after a certain time step, it is reset to $l_i=L$, where $L$ is the length-at-birth.
Then, a second cell with a new particle label and the same length $L$ is introduced and the total cell count $N$ is increased accordingly.
In the course of this cell division, the positions of the two daughter cells are shifted from the rod's original position by $L/2$ along the direction of the rod axis.

In the main text, the repulsion strength is fixed as $F_0=10^6 G/(\gamma L)$ and the length of the short axis of each rod is $d_0=L/8$.
The data presented in our plots are averaged over five simulation runs.

\subparagraph*{DDFT in one spatial dimension}

To illustrate the spatial evolution of the length distribution in DDFT, we consider a system in one spatial dimension described by the density $\rho(x,l,t)$.
As a minimal model, we take the growth function $G$ as a constant elongation rate \cite{you2018geometry,you2019monotomultilayer},
focus on a freely growing colony in the absence of an external potential $V_\text{ext}=0$
and employ a soft-repulsive pair interaction
\begin{equation}
 U(x-x',l,l') = k_\text{B}T\,\frac{2}{\sqrt{\pi}}\exp\!\left(-\frac{4(x-x')^2}{(l+l')^2}\right)
\label{eq_GaussianX}
\end{equation}
 in the form of Gaussian cores \cite{archer2001binary}
whose width follows from the lengths $l$ and $l'$ of two interacting bacteria.
Such a potential is conveniently included through the mean-field functional
\begin{equation}
\mathcal{F}_\text{ex}\left[\rho\right] = \frac{1}{2}\iiiint\mathrm{d}x\,\mathrm{d}x'\,\mathrm{d}l\,\mathrm{d}l'\,\rho(x,l)\, U(x-x',l,l')\,\rho(x',l')\,,
\label{eq_FexMF}
\end{equation}
where the integrals over $x$ and $x'$ run over the full one-dimensional space while $l$ and $l'$ are restricted under the present assumptions to the interval $[L,2L]$.
More general interactions could also be incorporated through appropriate alternative free-energy functionals~\cite{Evans1979,vanderlickpercus1989mixture,wittmann2017phase,wittmann2016fundamental}.

With the above choices, we rewrite
the length current from Eq.~\eqref{eq_DDFTlllterms} as
\begin{align}
J_l=  G\, \rho -D\, \frac{\partial \rho}{\partial l} - \frac{S}{k_\text{B}T}\,\rho\,\frac{\partial}{\partial l} {\frac{\delta\mathcal{F}_\text{ex}[\rho]}{\delta\rho}}\,.
\label{eq_current1d}
\end{align}
The one-dimensional version of Eq.~\eqref{eq_DDFTgcCONT_BC} then reads
\begin{align}
\frac{\partial \rho}{\partial t} &= D_x \frac{\partial^2 \rho}{\partial x^2} + \gamma^{-1}_x \frac{\partial}{\partial x}\biggl(\rho\,\frac{\partial}{\partial x} {\frac{\delta\mathcal{F}_\text{ex}[\rho]}{\delta\rho}}\biggr) \nonumber\\
&\quad - G \frac{\partial \rho}{\partial l} + D \frac{\partial^2 \rho}{\partial l^2} + \frac{S}{k_\text{B}T}\,\frac{\partial}{\partial l}\biggl(\rho\,\frac{\partial}{\partial l} {\frac{\delta\mathcal{F}_\text{ex}[\rho]}{\delta\rho}}\biggr)  \nonumber\\
&\quad  + \delta(l-L)J_l\!\left(x+\frac{L}{2},2L,t\right) \nonumber\\
&\quad  + \delta(l-L)J_l\!\left(x-\frac{L}{2},2L,t\right)\nonumber\\
&\quad - \delta(l-2L)J_l\left(x,2L,t\right)  \,,
\label{eq_DDFT1d}
\end{align}
where a detailed derivation is given in Supplementary Note 1 D.

Although the cell length $l$ can, in principle, take any positive value, we restrict it here to the fixed interval $l\in[L,2L]$  with constant length-at-birth $L$ and consider a directed boundary condition (see Supplementary Note 5 for a detailed description and also Eq.~\eqref{eq_directedboundary} below).
At $l=L$, this corresponds to a specific no-flux boundary condition in Eq.~\eqref{eq_current1d} imposed for the fluctuations (the term $\propto D$)
and to enforcing a vanishing density $\rho(x,l=L,t)=0$ if the drift-like terms (those $\propto G$ and $\propto S$) result in a negative contribution to the current.
At $l=2L$, we use an absorbing boundary.

\subparagraph*{Homogeneous size-resolved DDFT}

For the size-resolved calculations presented here, we employ a version of Eq.~\eqref{eq_DDFTgcCONT_BC} for the dynamics of $\rho(l,t)$ after averaging over positional and orientational coordinates, while assuming again that the external potential $V_\text{ext}(l,t)$ vanishes (or does not depend on the cell length) and taking a constant elongation rate $G$.
The resulting DDFT reads
\begin{align}
 \frac{\partial \rho}{\partial t}&=-\frac{\partial}{\partial l}J_l(l,t)\cr
 &\ \ \ \, +2\delta(l-L)J_l\left(2L,t\right)
 -\delta(l-2L)J_l\left(2L,t\right)\,,
 \label{eq_DDFTgcCONT_BC0d}
\end{align}
where the length current $J_l(l,t)$ is given by
\begin{align}
J_l=G\rho(l,t) -D \partial_l \rho(l,t) - S \rho(l,t)\bar{\rho}(t) \,.
\label{eq_Jll}
\end{align}
The third term of $J_l(l,t)$ is derived assuming a mean-field expression of $\mathcal{F}_\text{ex}[\rho]$, Eq.~\eqref{eq_FexMF},
 for soft interactions in the form of Gaussian cores, Eq.~\eqref{eq_GaussianX}.
  Hence, this size-resolved DDFT is fully consistent with Eq.~\eqref{eq_DDFT1d} and \eqref{eq_current1d} in one spatial dimension.
 The expression in Eq.~\eqref{eq_Jll} can also be used in any spatial dimension after absorbing a trivial dimensional scaling factor into $S$.

The spatially homogeneous nature of our size-resolved DDFT allows us to
consider two types of boundary conditions for $l\in[L,2L]$.
First, when working with Eq.~\eqref{eq_DDFTgcCONT_BC0d}, we use the directed boundary condition (compare Supplementary Note 5 A))
\begin{align}
D \left.\frac{\partial}{\partial l} \rho(l,t)\right|_{l=L} &\overset{!}{=} 0\,,\cr
\rho(l=L,t) &\overset{!}{=} 0 \ \ \text{if } \ S\bar\rho(t) > G \,,\cr
\rho(l>2L,t) &\overset{!}{=} 0\,
\label{eq_directedboundary}
\end{align}
for the current in Eq.~\eqref{eq_Jll}.
Second, a convenient alternative is to incorporate cell division through the oblique boundary condition
\begin{equation}
\rho(L,t)=2\rho(2L,t)\,,
 \label{eq_rhodivision1dHOM}
\end{equation}
assuming that individual cell growth is homeostatic, while rewriting Eq.~\eqref{eq_Jll} as
 \begin{align}
 \frac{\partial \rho(l,t)}{\partial t}=-\frac{\partial}{\partial l}J_l(l,t)\,,
 \label{eq_DDFTgcCONT_BC0dX}
\end{align}
which yields the DDFT equation
\begin{align}\!\!\!\!\!\!\!\!
\frac{\partial \rho}{\partial t}
&=\frac{\partial}{\partial l}
\left(-G\rho(l,t)+D \frac{\partial \rho(l,t)}{\partial l}
+ S \rho(l,t)\bar{\rho}(t) \right)\,. \!\!\!\!\!\!\!\!\!\!
 \label{eq_DDFT1dHOMexplicit}
\end{align}
This approximation allows for a detailed analytic understanding of the length-distribution dynamics,
where the effective elongation rate $G_\text{eff}(t)$ in Eq.~\eqref{eq_Geff} can be conveniently defined from the term on the right-hand side.
A more precise specification of $G_\text{eff}(t)$, a full derivation of Eq.~\eqref{eq_Jll}, details on the role of the different boundary conditions, analytic analysis and further results are provided in Supplementary Notes 2, 4 and 5.

\subparagraph*{DDFT for multiple bacterial species.}

It is in general straightforward to generalize a given DDFT model to mixtures of $\kappa$  species with different properties by adding an additional species label $i=1,\ldots,\kappa$
and consider the individual evolution equations for $\rho_i$, which are coupled via their collective interactions \cite{tevrugt2020revDDFT}.
For the purpose of the present study, we generalize the size-resolved DDFT from Eq.~\eqref{eq_DDFTgcCONT_BC0d} to
\begin{align}
 \!\!\!\!\!\!\!\!\!\frac{\partial \rho^{(i)}}{\partial t}&=-\frac{\partial}{\partial l}J_l^{(i)}(l,t)\cr
 &\ \ \ \, +2\delta(l-L)J_l^{(i)}\left(2L,t\right)
 -\delta(l-2L)J_l^{(i)}\left(2L,t\right)\
 \label{eq_DDFTgcCONT_BC0dmix}
\end{align}
with the currents
\begin{align}
\!\!\!J_l^{(i)}=G_i\rho^{(i)}(l,t) -D \partial_l \rho^{(i)}(l,t) - \sum_{j=1}^\kappa S_{ij} \rho^{(i)}(l,t)\bar{\rho}^{(j)}(t) \,.
\label{eq_Jllmix}
\end{align}
In these equations, we assumed for simplicity that all species have the same length-at-birth $L$ and the same magnitude $D$ of growth fluctuations.
Moreover, we consider $\kappa=2$ different species $A$ and $B$ and define the elongation rates $G_A:=G_1$ and $G_B:=G_2$, as well as the strengths $S_A:=S_{11}$ and $S_B:=S_{22}$ of the intra-species mechano-response, where we assume $S_{12}=S_{21}=(S_A+S_B)/2$ for the cross-interactions (see Supplementary Note 6 for further discussion).

\subparagraph*{Size-averaged logistic growth}

 The phenotemporal description of $\rho(l,t)$ in Eq.~\eqref{eq_DDFT1dHOMexplicit} represents an averaged model after integrating out positions and orientations of a more general DDFT for $\rho(\mathbf{r},\mathbf{p},l,t)$.
  In turn, if one is only interested in the increase of the density $\bar{\rho}(t)$ (or number of cells), we can show upon further averaging out the dependence of the cell size (see Supplementary Note 3) that our model is consistent with the logistic growth equation
 \begin{align}
 \partial_t\bar{\rho}(t)=\bar{\rho}(t)\left(R-\frac{S\ln2}{L}\bar{\rho}(t)\right),
  \label{eq_DDFTcg}
\end{align}
widely used to describe (space-resolved) population dynamics \cite{verhulst1838notice,vandermeer2010populations,korolev2010genetic,pigolotti2012population}.
From our Eq.~\eqref{eq_DDFT1dHOMexplicit} we identify here the overall growth rate  $R:=G\ln(2)/L+D(\ln(2)/L)^2$.
Solving Eq.~\eqref{eq_DDFTcg} with the initial condition $\bar{\rho}_0:=\bar{\rho}(0)$,
 we find an analytic expression for the time evolution of the total density $\bar{\rho}(t)$, as given by
Eq.~\eqref{eq_DDFTcgSOLrho}.
Note that generalized growth equations \cite{fujikawa2003new,pinto2022compressed} can also be derived within our framework from different microscopic interactions.

\subparagraph*{Experiments on growing bacterial colonies}

 We use non-motile strains of \textit{E.\ Coli} bacteria, NCM3722 delta-motA, growing at 30\textdegree{}C on a millimeter-thick agarose matrix.
The experimental time scales are short enough to ensure that the growing bacterial monolayers remain nutrient-replete and do not undergo physiological changes throughout the duration of the experiments.
Nutrient-limitation, if any, will impact all cells irrespective of their location in the colony.
This setup ensures that collective (mechanical) stresses constitute the main cause of (locally) limited cell growth reported in Fig.~\ref{figure1}a-h.

 All experiments have been performed for a minimum of three distinct biological replicates. Cells were grown and monitored using standard protocols and control experiments~\cite{dhar2022escape2d_time}.
 The growth of a single bacterium (or two initial cells, in some cases) into colonies was imaged while maintaining the growth temperature of 30\textdegree{}C within the microscope environment.
 Single bacteria acting as monoclonal nucleating sites expand horizontally on the nutrient-rich agarose layers.
 Initially, the colony expanded in two dimensions as a \textit{bacterial monolayer} over multiple generations, subsequently penetrating into the third dimension.

We visualize the colony growth over the entire period using time-lapse phase-contrast microscopy. For the current work, we focus primarily on the horizontal spreading of the colony and analyze the data till the transition to the multilayer structure sets off. Images were acquired using a Hamamatsu ORCA-Flash Camera ($1$ $\mu$m $= 10.55$ pixels) that was coupled to an inverted microscope (Olympus CellSense LS-IXplore). We use a 60X oil objective and, in some cases, 100X oil objectives to zoom into specific regions of the growing colonies.
Overall, this gave a minimum resolution of 0.11 $\mu$m.

Each experiment lasted typically 15h to 18h, allowing us to capture the mono-to-multilayer configurations and the structure and dynamics of multilayer colonies.
Prior to image acquisition, multiple locations on the agarose surface (where a single bacterium or up to two cells were present) were identified and recorded, allowing us to additionally extract technical replicates from the same sample.
The microscope was automated to scan these pre-recorded coordinates and to capture the images of the gradually increasing colonies after every five minutes while maintaining the focus across all the colonies captured.

We extracted the cell dimensions (width and length), position (centroid) and orientation of each bacterium from the phase-contrast images using the combination of open-source packages of Ilastik \cite{berg2019Ilastik} and ImageJ as well as MATLAB (MathWorks).
The combination of phase contrast and time-lapse imaging allowed us to quantify phenotypic traits at the resolution of individual cells and thereby extract the reported statistics after image analyses while ensuring that the cells do not tilt out of the plane \cite{beroz2018verticalization,nijjer2021mechanical} (see Supplementary Figure 8).
A detailed description of cell culturing, fabrication and imaging of cell monolayers can be found in Supplementary Note 7.
The obtained cell-length statistics are shown in Supplementary Tables 1-3 and provided as Supplementary Data 1-3.

\subparagraph*{Comparison of different approaches}

Our cell-based Langevin model essentially describes the same physics of growing bacterial colonies as our general DDFT.
However, the former conveniently ignores positional and rotational diffusion terms, which are typically required in DDFT.
Apart from this minor difference, Eqs.~\eqref{eq_LEr} and~\eqref{eq_LEt} are conceptually equivalent to the first and second terms in Eq.~\eqref{eq_DDFTgcCONT_BC}, respectively (the sole difference being the way interactions are chosen and implemented, as discussed below).
The dynamical description of cell size is the heart of our model and the four ingredients illustrated in Fig.~\ref{figure1}i are accounted for in all our approaches.
While the implementation of cell division is straightforward in the Langevin model, our DDFT requires the translation to a slightly different directed boundary condition~\eqref{eq_directedboundary}.
The description of growth and fluctuations in the Langevin model is stochastically equivalent to our DDFT when we compare the first two terms in Eq.~\eqref{eq_LangevinSI} to those of the probability current in Eq.~\eqref{eq_DDFTlllterms}, Eq.~\eqref{eq_current1d} and Eq.~\eqref{eq_Jll}, or to those on the right-hand-side of Eq.~\eqref{eq_DDFT1dHOMexplicit}.
Finally, despite the different appearance of the third term in these equations, the treatment of collective interactions is also conceptually equivalent: they are derived from the same expression as in configurational space (steric forces $\bm{F}_{ij}$ in the Langevin model and an excess free energy $\mathcal{F}_\text{ex}$ in DDFT), which is the defining feature of the mechano-response.

The specific choice and implementation of the interaction terms is what distinguishes our two computational approaches.
As we are interested in the basic phenomenology resulting from our model we have chosen to work with expressions that are standard in each case.
Hence, we implement our Langevin simulations with the established Hertzian overlap function~\eqref{eq_Hertz} \cite{allen2018bacterialreview,you2018geometry,you2019monotomultilayer} and equip our DDFT with a Gaussian soft-repulsive potential~\eqref{eq_GaussianX} which can be treated in the mean-field way~\eqref{eq_FexMF}.
To corroborate the general compatibility of both approaches, let us note that, for a well-mixed system, the DDFT equation from Eq.~\eqref{eq_DDFT1dHOMexplicit} is stochastically equivalent to the Langevin model
\begin{equation}
 \frac{\mathrm{d} l_i}{\mathrm{d}t}= G+ \sqrt{2D} \xi_i  - S\frac{N}{V}\,,
 \label{eq_LangevinSIxxx}
\end{equation}
where $N=\sum_i1$ is the current number of particles and $V$ is the total volume of the system, thus $\bar{\rho}=N/V$.
As shown in Supplementary Figure 4, the simulations of Eq.~\eqref{eq_LangevinSIxxx} with $l_i\in[L,2L]$ yield practically the same length distributions as DDFT with the directed boundary condition.

In view of the most accurate description of our experiments, a specific interaction potential would have to be measured for interacting cells and implemented in our mechanical terms.
However, our qualitative observations are largely independent of such a choice, as long as the assumed interaction is sufficiently repulsive.
The parameters entering our model equations are empirical and specific to the experimental nonequilibrium systems of interest.
In particular, the strength $S$ (or $\tilde{S}$) of the mechano-response is a measure of how strongly the growth behavior of a cell is actually affected by a mechanical stimulus - just like friction with the substrate determines the extent of the spatial drift induced by a repulsive force.
If desired, other growth-limiting effects that are not of mechanical origin (such as nutrient depletion) can be effectively described by an appropriate adaption of this parameter to experimental measurements.

 Although it is not the focus of the present work, we stress that the onset of
cell division can also be affected by different biological or mechanical mechanisms.
Hence, to better represent the real bacterial system, our basic model could be fine-tuned by allowing for a time-dependent length-of-birth $L(t)$ in future work.
For example, individual \textit{Escherichia coli} cells grow according to the adder model and divide after having grown by a certain length \cite{campos2014constant,taheri2015cell,si2019mechanistic}.
More specifically, taking a closer look at our experimental data, we find that the
periodicity of the oscillations in Fig.~\ref{figure4} decreases in the course of the colony evolution, i.e., on average, a cell divides every 33-36 minutes in the dilute case and every 24-27 minutes in the dense case. As the elongation rate, averaged over the colony, also decreases over time, this observation is accompanied by a reduction of the maximum length an individual cell reaches before the division event, i.e., the length-at-birth decreases from generation to generation.
 In addition, the length-at-birth depends on a cell's local position in relation to its neighbors in the growing colony.
Hence, we conclude that the observed decrease of the mean cell length in Fig.~\ref{figure4}
is consistent with our current model of a mechano-response depending on the local density (even in the simple form with a constant length-at-birth $L$) and that the individual growth kinetics play only a minor role.
In an extended model, the length-at-birth should thus also depend on the density, which could be modeled by similar terms as used for the length current in Eq.~\eqref{eq_DDFTlllterms}.

\subparagraph{Acknowledgments}

The authors would like to thank Michael te Vrugt, Nicola Pellicciotta, Marco Mazza, Simon Schnyder and Jens Elgeti for stimulating discussions and Marcel Funk for his contributions to implementing the size-resolved Langevin simulations~\eqref{eq_LangevinSIxxx}.
R.W.\ and H.L.\ acknowledge support by the Deutsche Forschungsgemein\-schaft (DFG) through the SPP 2265, under grant numbers WI 5527/1-1 (R.W.) and LO 418/25-1 (H.L.).
A.S.\ thanks the Institute for Advanced Studies, University of Luxembourg (AUDACITY Grant: IAS-20/CAMEOS), and the Luxembourg National Research Fund's ATTRACT Investigator Grant (Grant no.~A17/MS/11572821/MBRACE) and CORE Grant (C19/MS/13719464/TOPOFLUME/Sengupta) for supporting this work.\\

 \bibliographystyle{naturemag}

\end{document}